\documentclass[%
 reprint,
 amsmath,amssymb,
 aps,
]{revtex4-2}
\usepackage[utf8]{inputenc}

\usepackage{moreverb}
\usepackage[T1]{fontenc}
\usepackage{geometry}
\usepackage{float}
\usepackage{amsmath}
\usepackage{nameref}
\usepackage{algorithm}
\usepackage{amsthm}
\usepackage{amssymb}
\usepackage{caption}
\usepackage{subcaption}
\usepackage{comment}
\usepackage{fancyhdr}
\usepackage{graphicx}
\usepackage{bm}

\theoremstyle{example}

\newtheorem{theorem}{Theorem}

\theoremstyle{definition}
\newcommand{\grad}{\nabla}
\newcommand{\del}{\partial}
\newcommand{\half}{\frac{1}{2}}

\newcommand\BibTeX{{\rmfamily B\kern-.05em \textsc{i\kern-.025em b}\kern-.08em
T\kern-.1667em\lower.7ex\hbox{E}\kern-.125emX}}
\usepackage{graphicx}% Include figure files
\usepackage{dcolumn}% Align table columns on decimal point
\usepackage{bm}% bold math
\begin{document}

\preprint{APS/123-QED}

\title{Dimensional reduction of gradient-like stochastic systems with multiplicative noise via Fokker-Planck diffusion maps}% Force line breaks with \\

\author{Andrew Baumgartner}
 \email{andrew.baumgartner@isbscience.org}
\author{Sui Huang}%
 \email{sui.huang@isbscience.org}
\author{Jennifer Hadlock}
 \email{jennifer.hadlock@isbscience.org}
\author{Cory Funk}
 \email{cory.funk@isbscience.org}
 
\affiliation{%
Institute for Systems Biology \\
401 Terry Ave, Seattle, WA 98109
}%

\date{\today}% It is always \today, today,
%TC:ignore
\begin{abstract}
Dimensional reduction techniques have long been used to visualize the structure and geometry of high dimensional data. However, most widely used techniques are difficult to interpret due to nonlinearities and opaque optimization processes. Here we present a specific graph based construction for dimensionally reducing continuous stochastic systems with multiplicative noise moving under the influence of a potential. To achieve this, we present a specific graph construction which generates the Fokker-Planck equation of the stochastic system in the continuum limit. The eigenvectors and eigenvalues of the normalized graph Laplacian are used as a basis for the dimensional reduction and yield a low dimensional representation of the dynamics which can be used for downstream analysis such as spectral clustering. We focus on the use case of single cell RNA sequencing data and show how current diffusion map implementations popular in the single cell literature fit into this framework.
\end{abstract}
%TC:endignore

\maketitle
\section{Introduction}

The problem of dimensional reduction of single cell RNA sequencing data (sc-RNAseq) is a difficult one. The analysis of these high dimensional, discrete systems usually involves a panoply of preprocessing tools that map a vector $x \in \mathbb{N}_0^m$ (where $m$ is the total number of genes in the sample) to some dense subset of a submanifold embedded in the positive orthant of $\mathbb{R}^m$. A suite of tools is then used to ``learn" this data manifold and utilize geometric information to make claims about various biological processes, typically using $k$ nearest neighbor (kNN) graphs and/or clustering. The validity of this procedure has, and should be, questioned \cite{Gorin2022, Gorin2023} but nonetheless is motivated by the existence of a continuum limit of the more fundamental chemical master equations. In this work, we introduce a new twist on an old method that is meant to act as a bridge between this ``top-down" approach of manifold learning and the ``bottom-up" approach of chemical master equations via the spectra of the Fokker-Planck equation. \\

Specifically, we introduce a specific kNN graph whose normalized graph Laplacian generates the backward Fokker-Planck equation of generic, continuous model of single cell dynamics. The diffusion maps \cite{Coifman2006, Coifman2005} associated with this graph are then used for spectral clustering to identify cell types and subtypes. A general analysis of graph Laplacians was carried out in \cite{Ting2011}, where a correspondence was established between general, isotropic stochastic differential equations and graph Laplacians via drift-diffusion equations. By finding the correct normalization of the graph Laplacian, we can model the dynamics of a single cell moving through a potential landscape, a popular model for cellular development initially due to Waddington \cite{Huang2011, wang_quantifying_2011, Ferrell2012}. \\

Diffusion maps have been used by the sc-RNAseq community since at least 2015 \cite{Haghverdi2015}. They have primarily been used low dimensional representation of the \textit{geometry} of high dimensional data. This approach comes with many benefits, including being insensitive to the sampling density of the data. A more recent implementation introduced in \cite{Haghverdi2016}, and integrated into popular single cell packages such as scanpy \cite{Wolf2018} and Seurat \cite{Stuart2019}, introduced local bandwidths and weights to the graph normalization for better visualization, but no longer carries a direct interpretation to the geometry of the data manifold, as we will see below. \\

In section II we will summarize prior results and explain how we used them in our specific graph construction. In section III we derive a general formula for the potential of the system in terms of nearest neighbors, and show how this simplifies when one approximates the strength of the noise as a linear function of the distance to the $k^{th}$ nearest neighbor. We introduce our choice of kernel to obtain the Fokker-Planck diffusion maps in section IV and apply it to numerous simulated and real biological systems in section V. We conclude with a discussion in section VI. For simplicity, we assume $x, y \in \mathbb{R}^m$ and omit bold faced type for vectors.  Additionally, we assume the existence of an optimal $k$ such that the kNN graph yields an appropriate and accurate discretizaiton of the data manifold. In reality, many $k$'s should be used to track any results across many scales.

\section{The circle of diffusion processes}
The following relies on a result concerning general diffusion processes. These are defined by the solution to the following stochastic differential equation
\begin{equation} \label{eq:driftDiff}
dx = \mu(x)dt + \sigma(x) dW
\end{equation}
where $dW$ is a Wiener process. The arbitrary functions $\mu(x)$ and $\sigma(x)$ are the drift and diffusion terms and are given by vectors and matrices, respectively, which depend on the state. The generator $\mathcal{G}$ of this diffusion processes is given by
\begin{equation} \label{eq:elliptic}
\mathcal{G}f = \frac{1}{2} \sum_{i,j,k} \sigma(x)_{ik}\sigma(x)_{kj}^T \frac{\partial^2f}{\partial x_i \partial x_j} + \sum_i \mu_i(x)\frac{\partial f}{\del x_i} .
\end{equation}
Thus, for every diffusion process there is a corresponding elliptic equation which generates the process. What follows is a result from \cite{Ting2011} that establishes a connection between eq. \eqref{eq:elliptic} and normalized graph Laplacians. The graph construction can be viewed as a discrete approximation to eq. \eqref{eq:elliptic}. \\

We assume we have a translationally and rotationally invariant kernel $K_0(|x-y|)$ with compact support. We then introduce a weight function $w_x(y) \ge 0 $ and a bandwidth function $r_x(y) >0 $ which admit a Taylor expansion about $y=x$ with bounded error, denoted by $w^{(n)}_x(y)$ and $r^{(n)}_x(y)$. We then rescale the kernel:
\begin{equation} \label{eq:kern}
K(x,y) = w^{(n)}_x(y) K_0 \left( \frac{|x-y|}{h_n r^{(n)}_x(y)} \right)
\end{equation}
where $h_n > 0$ is a bandwidth scaling which goes to 0 in the continuum limit. The following theorem can be found in \cite{Ting2011}:
\begin{theorem}
\cite{Ting2011} Assume the standard assumptions for the diffusion approximation to hold eventually with probability 1. If the bandwidth scaling $h_n$ satisfies $h_n \to 0$ and $nh_n^{m+2}/log(n) \to \infty$ then for graphs constructed using the kernel 

\begin{equation*} 
K(x,y) = w^{(n)}_x(y) K_0 \left( \frac{|x-y|}{h_n r^{(n)}_x(y)} \right)
\end{equation*}

there exists a constant $Z_{K_0,m}>0$ depending only on the base kernel $K_0$ and the dimension of the data manifold $m$ such that for $c_n = Z_{K_0,m}/h^2$, 

\begin{equation*}
    -c_n L^{(n)}_{rw}f \to Af
\end{equation*}

where A is the infinitesimal generator of a diffusion process with the following drift and diffusion terms given in normal coordinates
\begin{align} \label{eq:drift+diff}
 \begin{split}
 \mu(x) &= r_x(x)^2 \left(\grad_y \ln p(y) + \grad_y \ln w_x(y) \right. \\
       & \left. \quad + (m+2)\grad_y r_x(y) \vphantom{\grad_y \ln p(y)}\, \right)_{|y=x}  
 \end{split} \\ 
 \sigma(x)\sigma(x)^T &= r_x(x)^2 I
\end{align}
where all derivatives are taken with respect to $y$ (in appropriate coordinates) and evaluated at $y=x$, $p$ is the local sampling density at $x$ and $I$ is the $m\times m$ identity matrix. 
\end{theorem}
This allows us to approximate any isotropic equation of the form \eqref{eq:elliptic} via a graph with edge weights given by equation \eqref{eq:kern} with a clever choice of bandwidth and weight functions. \\

The original authors of the diffusion maps in \cite{Coifman2005, Coifman2006, Coifman2008} introduced the following normalization: 
\begin{equation}\label{eq:diffmapOG}
w_x(y) = p(x)^{-\alpha} p(y)^{-\alpha}, \,\,\, r_x(y) = 1
\end{equation}
which leads to
\begin{equation}
\mathcal{G}f = \frac{1}{2}\grad^2 f + (1-\alpha)\grad (\ln p) \cdot \grad f.
\end{equation}
When choosing $\alpha=1$ we recover the Laplacian of the manifold $\grad^2$ and the $p$ dependence drops out. This normalization was used in \cite{Haghverdi2015} for single cell data. A locally scaled version was introduced in \cite{Haghverdi2016} that induces a spurious drift term.

\section{Noise induced drift and the equilbrium distribtuion}
The work presented in \cite{Coifman2005, Coifman2006, Coifman2008} offers a way to represent gradient-like stochastic systems with additive noise via diffusion maps. This is done by taking $\alpha = 1/2$ in equation \eqref{eq:diffmapOG} and assuming the local density is of Boltzmann form, i.e. $p=e^{-U}$ for some potential $U$. This results in the following drift-diffusion equation 
\begin{equation} \label{eq:bFPE}
-\frac{\del f}{\del t} =  \grad^2 f + \grad U \cdot \grad f
\end{equation}
which is the backwards Fokker-Planck equation of the system
\begin{equation*}
dx = -\grad U dt + \sqrt{2} dW.
\end{equation*}
Solutions to \eqref{eq:bFPE} can be approximated by an eigenvalue expansion using the eigenvalues and eigenvectors of the graph Laplacian. The diffusion map gives a low dimensional representation of the system allowing for the identification of local minima and the paths connecting them. Moreover, this is the optimal dimensional reduction technique with respect to the standard error between the truncated eigenvalue expansion and the true solution \cite{Coifman2008}. This choice of local density is equivalent to saying that the system is in local ``information-thermodynamic" equilibrium with respect to the potential $U$. \\

Extending this procedure to systems with multiplicative noise is not so straightforward due to an ambiguity in the integration procedure of the associated stochastic differential equation eq. \eqref{eq:driftDiff}. The ambiguity is most easily seen through the corresponding Langevin equation
\begin{equation} \label{eq:lang}
\frac{dx}{dt} = \mu(x) + \sigma(x)\eta(t)
\end{equation}
where $\eta(t) \sim dW/dt$. This is the source of the Ito vs. Stratonovich controversy. The choice does not matter in systems with additive noise as well as systems with internal noise. The only situation in which the choice has physical consequences is in systems with external noise, whereby the Stratonovich interpretation is the correct one \cite{vanKampen1981}. scRNAseq data is unique in that it has both internal and external noise. Common preprocessing pipelines are meant to eliminate technical (external) noise and only keep biological (internal) noise, although their efficacy is widely debated (See \cite{Gorin2022, Gorin2023} for alternative approaches). \\

In the Stratonovich picture, the Fokker-Planck equation with $\mu = -\grad U$ becomes 
\begin{equation}
\frac{\del p}{\del t}=-\grad\cdot\left(\left(-\grad U+\half\grad\sigma^{2}\right)p-\half\grad\left(\sigma^{2}p\right)\right)
\end{equation}
which leads to an equilibrium distribution of the form 
\begin{equation} \label{eq:equil}
p=\frac{\mathcal{N}}{\sigma^{2}}\exp\left\{ \int^{x}\frac{2\left(-\grad U+\half\grad\sigma^{2}\right)}{\sigma^{2}}dz\right\} 
\end{equation}
where $\mathcal{N}$ is a normalization factor. The term $\half \grad \sigma^2$ is a ``noise induced drift" term. The consequences of this term on the Waddington landscape was beautifully illustrated in \cite{Coomer2022}. There, the authors show that this term flattens out existing minima of the potential and changes their location \cite{Coomer2022}. We could proceed by defining an effective potential as $U_{eff} = -\ln p$, but this obscures the physics of the underlying stochastic process. Luckily, when we take the noise term to be proportional to the $k^{th}$ nearest neighbor of $x$, things simplify and we can approximate the potential using only knowledge of the nearest neighbors. \\

To achieve this we utilize a trick from Ting et. al \cite{Ting2011} and use the kNN density estimate \cite{Loftsgaarden1965} as our local density. In $\mathbb{R}^m$ we have
\begin{equation}
p_{kNN}(x) = \frac{k}{N}\frac{1}{\Omega_m \rho(x)^m}.
\end{equation}
where $\Omega_m$ is the volume of the unit $m$-ball, and $\rho(x)$ is the distance to the $k^{th}$ nearest neighbor. If we equate this to the local equilibrium density with the noise induced drift term, we can solve for the potential in terms of the $k$-nearest neighbor field and the strength of the noise. To do so, we start with 
\begin{equation}
\frac{\mathcal{N}}{\sigma^{2}}\exp\left\{ \int^{x}\frac{2\left(-\grad U+\half\grad\sigma^{2}\right)}{\sigma^{2}} \cdot dz\right\} \approx\frac{C}{\rho(x)^{m}}
\end{equation}
where $C=k/(N\Omega_m)$ and the integral is shorthand for a line integral measured from the zero point of the potential, i.e. from $x_0$ to $x$ such that $U(x_0) = 0$. After a little algebra which we will not recapitulate here, we wind up with the general relation
\begin{equation} \label{eq:generlU}
U(x) =\int^{x}\sigma^{2}\grad\ln\left(\rho^{m/2}\right)dz. 
\end{equation}
If we then take $\sigma = \alpha \rho$ for $\alpha \in (0,1]$, this simplifies further to 
\begin{equation}
    U(x)=\frac{m\alpha^2}{4}\rho(x)^2.
\end{equation}
Thus, in order to approximate the SDE
\begin{equation}
    dX = -\grad U dt + \sigma(x)dW
\end{equation}
we need a normalization which gives 
\begin{equation} \label{eq:targ}
dX = -\frac{m\alpha^2}{4}\grad \rho(x)^2dt + \alpha\rho(x)dW
\end{equation}
up to some scale factor. 

\section{Fokker-Planck diffusion maps}
There are many different choices of kernel that will converge to the same equation in the continuum limit. The results of Ting et al. \cite{Ting2011} give an infinite family of graph normalizaitons which lead to the same drift-diffusion equation. This means that the identification of long-time macroscopic dynamics will be the same for these different graphs, while the short-time, microscopic dynamics will be different. For downstream applications this means that things like boundaries between detected clusters on the graph will be different between graphs with different kernels, while the macroscopic, coarse grained clusters will all converge. This suggests shying away from graph-based clustering in favor of spectral clustering. \\

We use a simple symmetric Gaussian kernel constructed as follows: start with a base kernel of the form 
\begin{equation}\label{eq:kernUse}
K(x,z) = e^{-(x-z)^2/\rho(x)^2}
\end{equation}
and symmetrize it. For instance, one can do the following:
\begin{align} \label{eq:kernUse}
    K_{sym}(x,y) &= \int K(x,z)K(y,z)dz  \\ 
     &= \left( \frac{\pi \rho(x)^2\rho(y)^2} {\rho(x)^2+\rho(y)^2} \right)^{m/2} e^{-(x-y)^2 / (\rho(x)^2+\rho(y)^2)}.
\end{align}
In discrete notation, this is equivalent to 
\begin{equation}
K_{sym} = KK^T.
\end{equation}
This normalization leads to the following stochastic differential equation:
\begin{equation}
dX = (1-\frac{m}{4})\grad \rho(x)^2 + \rho(x)dW
\end{equation}
which for $m \gg 4$ agrees with \eqref{eq:targ}. The downside to this symmetrization is that it effectively turns the nearest neighbor graph into a shared nearest neighbor graph, and so our graph may not be fully connected. To see this, note that the initial nearest neighbor kernel can be extended to all data points by including a Heaviside step function as in \cite{Ting2011}:
\begin{equation}
    K_{full}(x,z) =K(x,z)\Theta(|z-x|<\rho(x)).
\end{equation}
When we perform the integral in eq. to symmetrize \eqref{eq:kernUse} we obtain
\begin{equation}
 \begin{split}
    &K_{sym}(x,y) = \\
    &\int K(x,z)K(y,z)\Theta(|z-x|<\rho(x))\Theta(|z-y|<\rho(y)) dz. 
 \end{split}
\end{equation}
which is zero unless the neighborhoods of $x$ and $y$ are overlapping. Another option which is guaranteed to return a connected graph is to take $K$ as:
\begin{equation}
K_{sym}(x,y) = K(x,y) + K(y,x)
\end{equation}
or, in matrix notation
\begin{equation}
    K_{sym} = K + K^T.
\end{equation}
The linearity of the graph Laplacian and Taylor expansions used in \cite{Ting2011} allows us to add the contributions to eq. \eqref{eq:drift+diff} from both terms separately. Doing so yields the following SDE:
\begin{equation}
dX = \left(1-\frac{m}{2} \right)\grad\rho^2 dt + \sqrt{2}\rho(x)dW.
\end{equation}
We can redefine $\rho \to \rho/\sqrt{2}$ to instead obtain 
\begin{equation} \label{eq:final}
dX = \half \left(1-\frac{m}{2} \right)\grad\rho^2 dt + \rho(x)dW.
\end{equation}
which for $m\gg2$ reduces to the correct result. The eigenvectors and eigenvalues of the normalized graph Laplacians of these graphs can be used as a basis for dimensional reduction just like the diffusion maps of \cite{Coifman2005}. They can be used for visualizing the potential, as a basis for spectral clustering, or for finding low dimensional dynamical systems which approximate the original system as in \cite{Coifman2008,Nadler2006}.  \\

\begin{comment}
The symmetrization step is necessary for the validity of this approach. Symmetric Markov chains such as the ones we are constructing correspond to reversible processes. In using the backwards Fokker-Planck equation for the basis of our dimensional reduction, we are implicitly assuming our system to be reversible. Ultimately our interpretation of this method stems from the fact that the forwrd and backward Fokker-Planck equations are adjoints on the space of square integral functions, $L^2(\mathcal{F})$. As such, they share a set of eigenvalues and their eigenvectors are related to one another by a simple normalization. Specifically, the forward Fokker-Planck equation tells us about the forward time evolution of the probability density starting from some initial point at time $t=t_0$, while the backward Fokker-Planck equation tells us how the probability evolves backwards in time as a function of the initial conditions. Thus, equivalence of these descriptions requires the system to be reversible, so that forward and backward time evolution 
\end{comment}

The local bandwidth and weight functions for diffusion maps introduced in \cite{Haghverdi2016} can also be interpreted in this framework, although the initial justification for their construction is unclear. The authors model a cell in gene expression space as a Gaussian blob using the following: 
\begin{equation}\label{eq:diffmap2016}
Y_{x}(x')=\left(\frac{2}{\pi\sigma_{x}^{2}}\right)^{1/4}e^{-\frac{|x-x'|^{2}}{\sigma_{x}^{2}}}.
\end{equation}
We believe this normalization is unjustified, since for $x, x' \in \mathbb{R}^m$ the integral over $x'$ is
\begin{equation}
\begin{split}
    \int d^{m}x'&\left(\frac{2}{\pi\sigma_{x}^{2}}\right)^{1/4}e^{-\frac{|x-x'|^{2}}{\sigma_{x}^{2}}} \\
    &=\left(\frac{2}{\pi\sigma_{x}^{2}}\right)^{1/4}\left(\pi\sigma_{x}^{2}\right)^{m/2}\\
&=2^{\frac{1}{4}}\pi^{\frac{2m-1}{4}}\sigma_{x}^{m-\half}
\end{split}
\end{equation}
which does not have any clear biological or biophysical interpretation. Moreover, when constructing the graph kernel as as the overlap of these Gaussian blobs, the authors of \cite{Haghverdi2016} use
\begin{equation}
\begin{split}
K(x,y)&=\int dzY_{x}(z)Y_{y}(z) \\
&=\sqrt{\frac{2\sigma_{x}\sigma_{y}}{\sigma_{x}^{2}+\sigma_{y}^{2}}}e^{-\frac{|x-y|^{2}}{2(\sigma_{x}^{2}+\sigma_{y}^{2})}}
\end{split}
\end{equation}
which is incorrect for $x,y,z\in \mathbb{R}^m$. The correct result should be
\begin{equation}
\begin{split}
K(x,y)&=\int dzY_{x}(z)Y_{y}(z) \\
&=\frac{4^{1/4}\pi^{(m-1)/2}\left(\sigma_{y}^{2}\sigma_{y}^{2}\right)^{(m-2)/2}}{(\sigma_{x}^{2}+\sigma_{y}^{2})^{m/2}}e^{-\frac{|x-y|^{2}}{(\sigma_{x}^{2}+\sigma_{y}^{2})}}.
\end{split}
\end{equation}
which reduces to their result if $m=1$. Following \cite{Coifman2005}, the authors then define $Z(x) = \sum_y K(x,y)$ so the final graph kernel is 
\begin{equation}
W(x,y) = Z(x)^{-1}Z(y)^{-1}K(x,y).
\end{equation}
The definition of $Z(x)$ employed here is the kernel density estimate of $p(x)$ using the locally scaled kernel This follows from \cite{Coifman2005}. \\

Biological interpretation and incorrect integrals aside, the normalization of \cite{Haghverdi2016} et al. can still be interpreted within the framework of Theorem 1. A little algebra shows that these weight and bandwidth functions yield a graph which models the following SDE:
\begin{equation}
dX=(m+\frac{3}{2})\grad\sigma_{x}^{2}dt+2\sigma_{x}dW. 
\end{equation}
Upon taking $\sigma = \rho(x)/2$ we obtain
\begin{equation}\label{eq:hagSDE}
dX=\frac{1}{4}(m+\frac{3}{2})\grad\rho^{2}dt+\rho(x)dW. 
\end{equation}
Crucially the drift terms in eqs. \eqref{eq:hagSDE} and \eqref{eq:final} differ by a minus sign. In the Waddington landscape view of cell development, cell types and subtypes correspond to local minima of the potential. In analogy to real, physical potentials, the difference in minus signs determines whether the force experienced by a given cell will drive them towards a minimum or a maximum. These corresponds to situations whereby $\mu(x) = -\grad U$ versus $\mu(x) = \grad U$ respectively. This has the consequence that the SDE given in \eqref{eq:final} will drive cells towards more dense regions (those with $\grad \rho < 0$), while the SDE in \eqref{eq:hagSDE} will drive cells towards less dense regions (those with $\grad \rho >0$). The latter is good for visualization, since it will ``spread out" the data, so to speak, while the former is more appropriate for clustering and identification of cell types, since it drives cells towards their closest minimum. 

\section{Illustrations and applications}
We now turn to specific examples to illustrate our proposed diffusion map. We first consider the dynamics of two multidimensional Ornstein-Uhlenbeck processes--one with additive noise and one with linear, multiplicative noise-- and show how our Fokker-Planck diffusion maps can reflect the dynamics of a single trajectory in the system. Next, we turn to a multidimensional double well potential with orthogonal noise directions and demonstrate how an ensemble of trajectories can induce a clear separation of minima in the resulting diffusion map. Finally, we look at the PBMC3k set and compare to the implementation of Haghverdi et. al. \cite{Haghverdi2016} to demonstrate the differences derived above. 

\subsection{Ornstein-Uhlenbeck processes} 
%TC:ignore

\begin{figure}
     \begin{subfigure}{0.3\textwidth}
         \includegraphics[scale = .4]{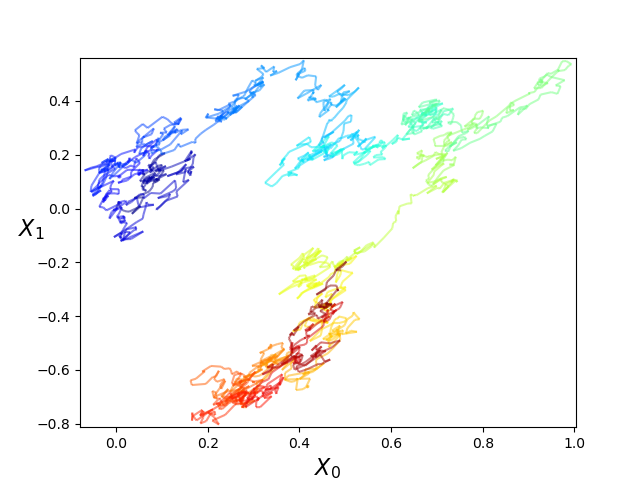}
         \centering
         \caption{2D projection of our 10D Ornstein-Uhlenbeck (O-U) trajectory with additive noise from t=0 (blue) to t=1 (red)}
         \label{fig:OUaddTraj}
     \end{subfigure}\\
     \begin{subfigure}[b]{0.3\textwidth}
         \includegraphics[scale = .25]{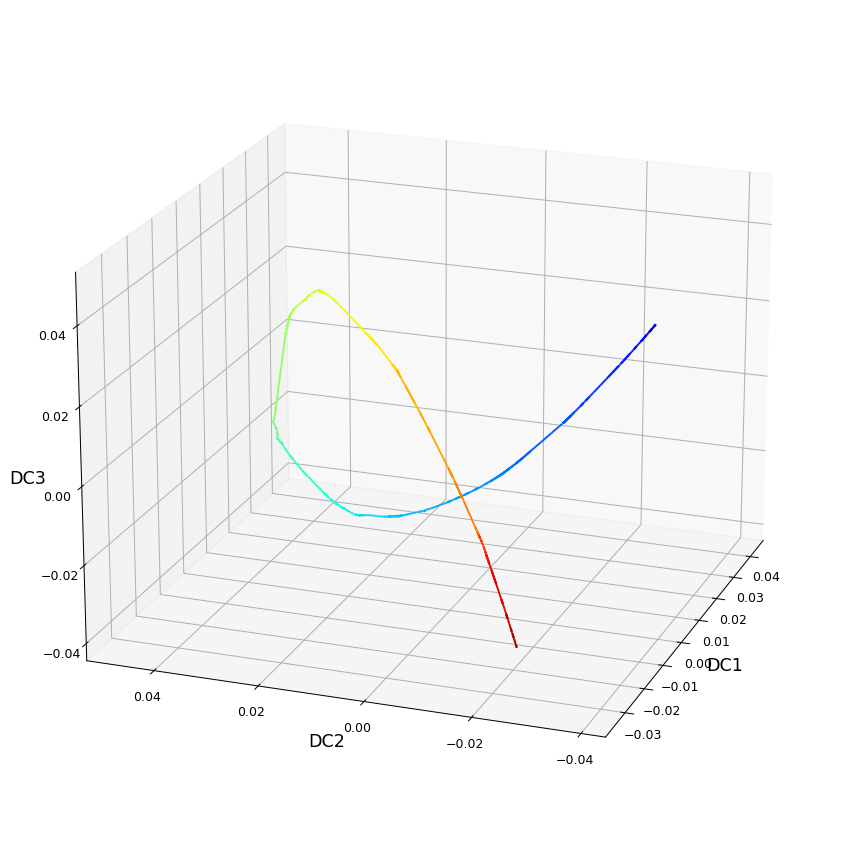}
         \centering
         \caption{The first three diffusion coordinates of the O-U trajectory.}
         \label{fig:addDiffMap}
     \end{subfigure}  \\
     \caption{A single trajectory for an Ornstein-Uhlenbeck process in 10D with additive noise and the corresponding diffusion map. The diffusion coordinates correspond to products of Hermite polynomials as shown in \cite{Coifman2005}. The matrices $\alpha$ and $\beta$ have random entries sampled from a uniform distribution over $[0,1]$}
     \label{fig1}
\end{figure}
 
\begin{figure}
     \begin{subfigure}{0.3\textwidth}
         \includegraphics[scale = .4]{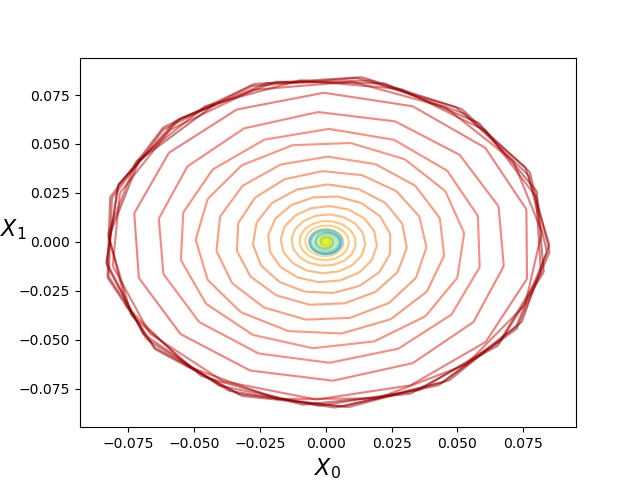}
         \centering
         \caption{2D projection of our 10D Ornstein-Uhlenbeck (O-U) trajectory with linear, multiplicative noise from t=0 (blue) to t=1 (red)}
         \label{fig:OUmultTraj}
     \end{subfigure}\\
     \begin{subfigure}[b]{0.3\textwidth}
         \includegraphics[scale = .25]{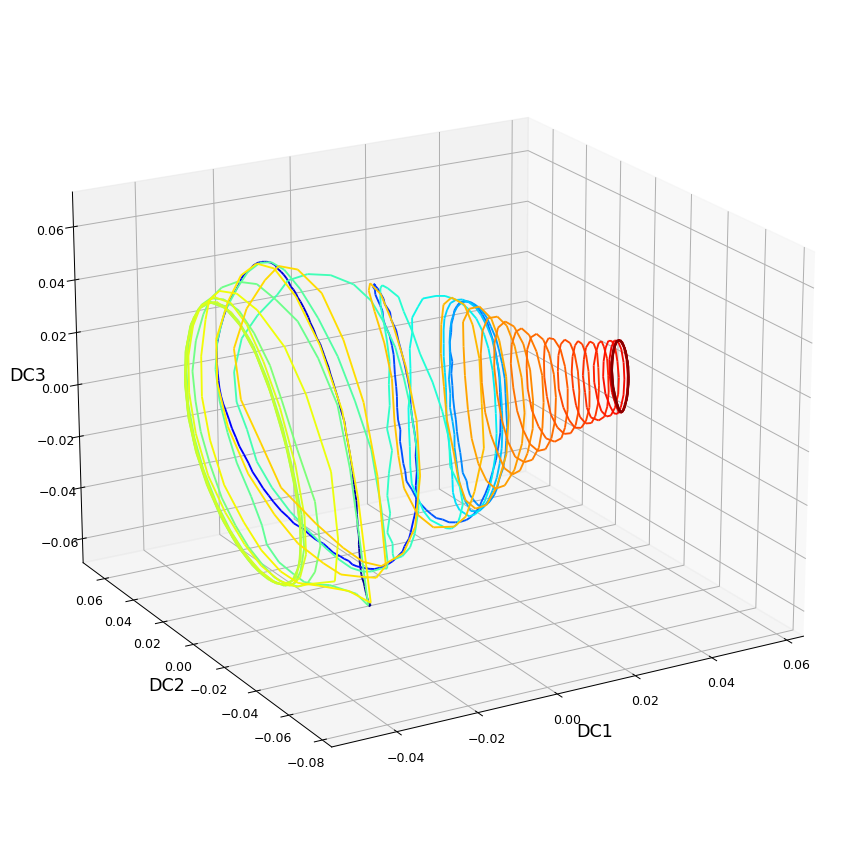}
         \centering
         \caption{The first three diffusion coordinates of the O-U trajectory.}
         \label{fig:addDiffMap}
     \end{subfigure}  \\
     \caption{A single trajectory for an Ornstein-Uhlenbeck process in 10D with linear, multiplicative noise with $\omega = .05$ and $\beta = .001$. }
     \label{fig2}
\end{figure}

\begin{figure}[t!]
     \begin{subfigure}{0.3\textwidth}
         \includegraphics[scale = .45]{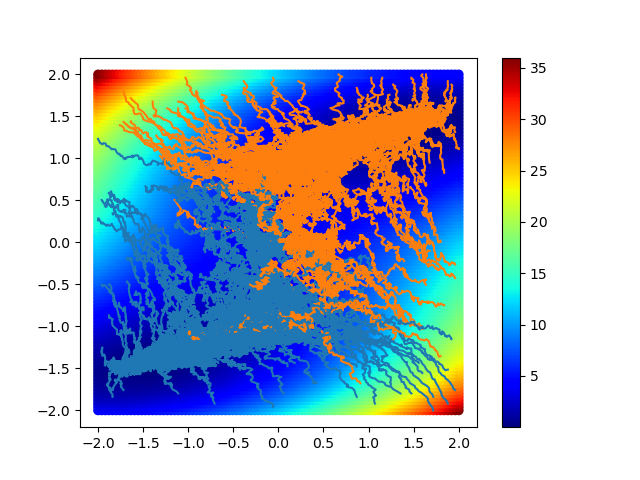}
         \centering
         \caption{Ensemble of trajectories of the 10D double well. The potential is localized in 2D and orthogonal directions are subject to pure diffusion. The colorbar denotes the value of the potential.  }
         \label{fig:doubleWellTraj}
     \end{subfigure}\\
     \begin{subfigure}[b]{0.3\textwidth}
         \includegraphics[scale = .23]{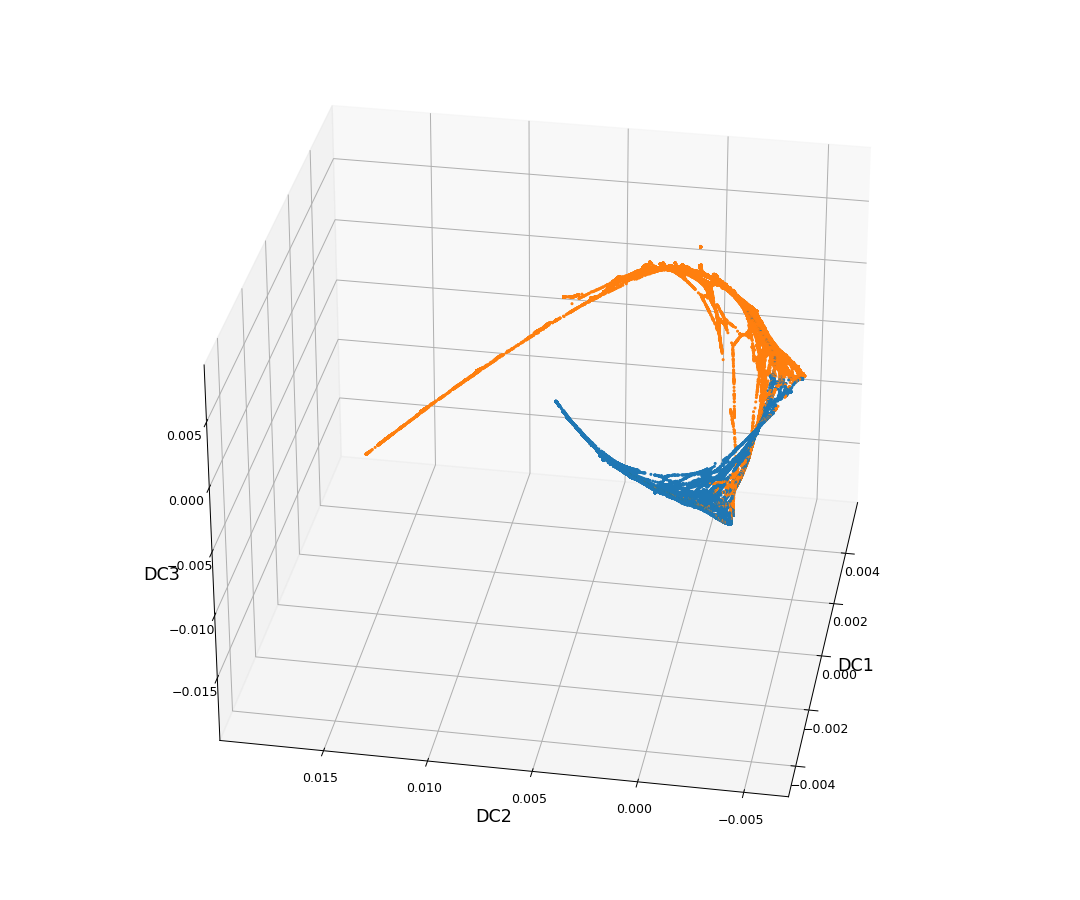}
         \centering
         \caption{The first three diffusion coordinates of our 10D ensemble of trajectories, showing separation between the two minima of the potential.}
         \label{fig:doubleWellDiffMap}
     \end{subfigure}  \\
     \caption{Our double well in eq. \eqref{eq:double_well} with $d=2$, $a=1$, $b=c=4$ and $\sigma(x) = 1-.2x^2$. Trajectories are colored by the minima they equilibrium to. Note the presence of trajectories which start in one minima and end up in another in both (a) and (b). }
     \label{fig3}
\end{figure}

\begin{figure}[t!]
     \begin{subfigure}{0.3\textwidth}
         \includegraphics[scale = .3]{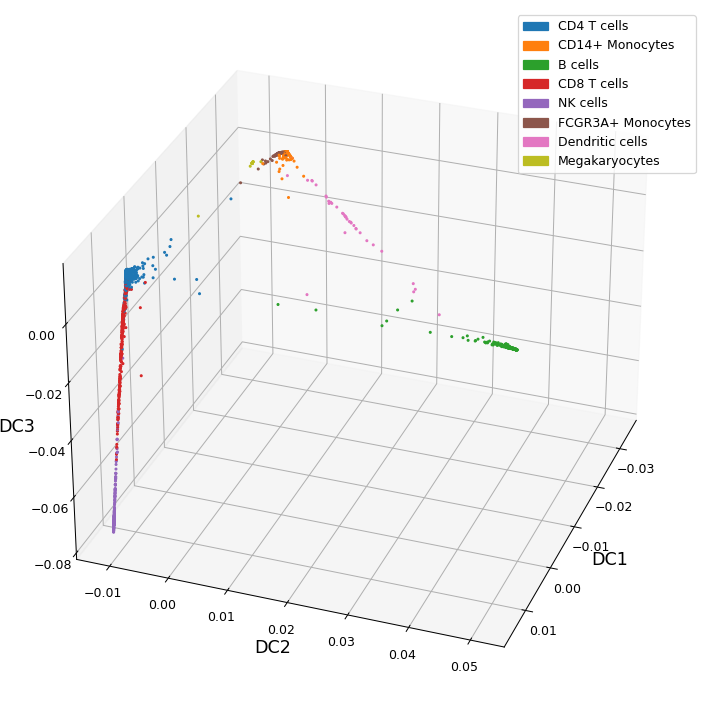}
         \centering
         \caption{The first three diffusion coordinates using the Fokker-Planck diffusion maps.}
         \label{fig:diffmapMe}
     \end{subfigure}\\
     \begin{subfigure}[b]{0.3\textwidth}
         \includegraphics[scale = .3]{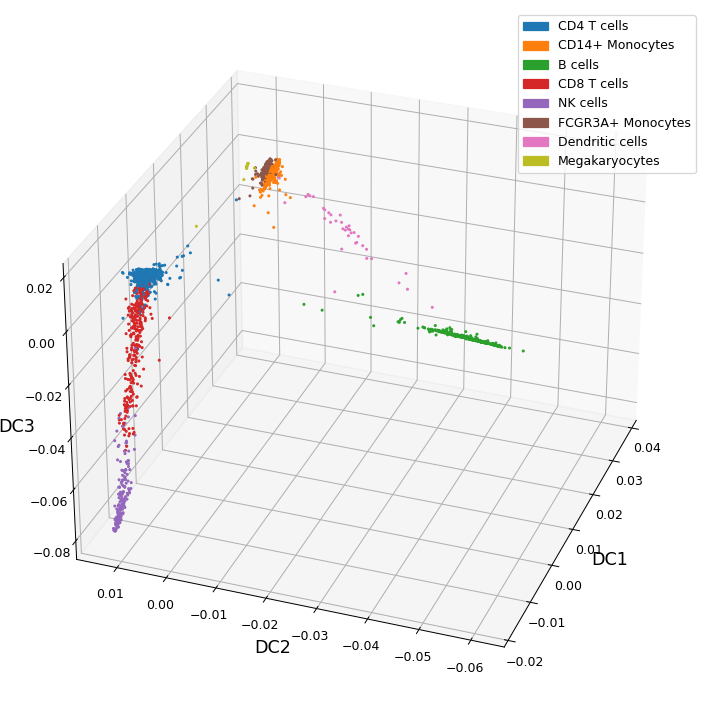}
         \centering
         \caption{The first three diffusion coordinates using the scanpy's diffusion maps.}
         \label{fig:diffMapscanpy}
     \end{subfigure}  \\
     \caption{The first three diffusion coordinates using the two different normalizations for diffusion maps discuss. Note the FP diffusion maps tend to clump similar cells together, while the scanpt diffusion maps spread them apart. This is a consequence of the relative sign in the force term of the corresponding SDE.}
     \label{fig4}
\end{figure}

\begin{figure}[t!]
     \begin{subfigure}{0.3\textwidth}
         \includegraphics[scale = .4]{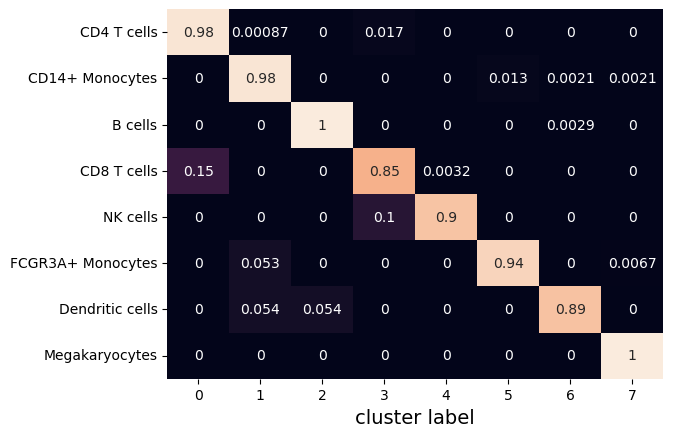}
         \centering
         \caption{Confusion matrix for hierarchical clustering on FP diffusion maps}
         \label{fig:confMate}
     \end{subfigure}\\
     \begin{subfigure}[b]{0.3\textwidth}
         \includegraphics[scale = .4]{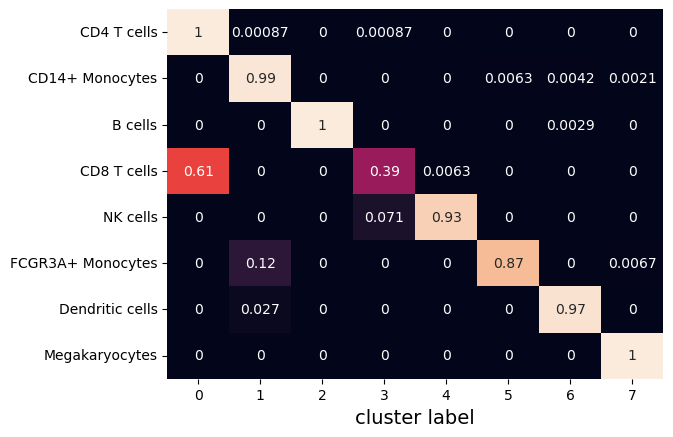}
         \centering
         \caption{Confusion matrix for hierarchical clustering on scanpy diffusion maps}
         \label{fig:cm1}
     \end{subfigure}  \\
     \caption{Confusion matrices for hierarchical clustering on the PBMC3k data set. Clustering was performed on the first 6 nontrivial diffusion coordinates. This corresponds to the number of eigenvalues before the largest spectral gap. }
     \label{fig5}
\end{figure}

%TC:endignore
The Ornstein-Uhleneck (O-U) process with additive noise is the rare example of a stochastic system where both the stochastic differential equation and the diffusion map can be solved exactly making it a perfect sanity check for our new normalization. Specifically, we have
\begin{equation}
    dx = -\alpha (x-\gamma) + \beta dW
\end{equation}
where, in general, $\alpha$ and $\beta$ are matrices. This system was treated in \cite{Coifman2005} where it was shown that the diffusion maps can be written in terms of (products of) Hermite polynomials. Figure \ref{fig1} illustrates these diffusion maps on a single trajectory in 10D. \\

Next we look at an O-U process linear multiplicative noise. Specifically, we consider 
\begin{equation}
dx=F x dt + G x dW.
\end{equation}
This can be solved if the matrices $F$ and $G$ commute, giving  
\begin{equation}
x(t) = e^{(F-\frac{1}{2}G^2)t + GW(t)}x(0).
\end{equation}
In particular, we use the following matrices
\begin{align} \label{eq:FandG}
    F & = \text{antidiag}(\omega, -\omega ,\omega,\dots,-\omega) \\
    G &= \text{diag}(\beta,\dots,\beta).
\end{align}
The results are shown in figure \ref{fig2}. 

\subsection{Multidimensional double well}
Next we consider a double well potential in multiple dimensions subject to multiplicative noise. We use the following form of the potential

\begin{equation} \label{eq:double_well}
U(x) =x_{0}^{2}(ax_{0}^{2}-b)+\frac{b^{2}}{4a}+\half\sum_{i=0}^{d-2}c_{i}(x_{i}-x_{i+1})^{2}
\end{equation}
for some $d<D$ where $D$ is the dimensional of the system. This localizes the potential into a $d$-dimensional subspace of the full phase space with pure diffusion in the orthogonal directions. This particular form is meant to mimic a system in which the data lives on some low dimensional subspace of the entire phase space, with noise in orthogonal directions. We simulate multiple trajectories in this space and run our diffusion maps on the resulting data set. Our results are shown in figure \ref{fig3}, where we choose 
\begin{equation}
    \sigma(x) = 1-.2 x^2.
\end{equation}
We see a clear separation of trajectories falling into their distinct minima with interpolation between the two representing trajectories which jump the potential barrier, suggesting the diffusion maps recapitulate the dynamics of the entire ensemble.

\subsection{Single Cell RNAseq--PBMC3k}

Finally we look at our primary use case--dimensional reduction and subsequent clustering of scRNAseq data. We compare our results to the current implementation of diffusion maps in scanpy \cite{Wolf2018}, with the method laid out in \cite{Haghverdi2016}. The diffusion map for both of these is shown in figure \ref{fig4}. We see that our implementation tends to group cells closer together, while the scanpy implementation spreads them out more. This is due to the relative minus sign in the force terms. \\

We then use these diffusion maps as a basis for hierarchical clustering. We restrict our diffusion maps to be above the first spectral gap (corresponding to the first 6 nontrivial diffusion coordinates), perform hierarchical clustering and compare our clusters to the labels of the PBMC3k set. Specifically, we used the processed data available through the scanpy API. Our results are shown in figure \ref{fig5}. Note that overall the two methods are comparable, with only a $2\%$ difference in the balanced accuracy. However, we find that our method is preferable when cells are more difficult to distinguish, as in the case of CD4 and CD8 T-cells, and the CD14 and FCGR3A+ monocytes. We believe this is due to the sign of our potential term which, as discussed above, drive cells to the nearest minima, as opposed to the scanpy implementation which drives cells away from the minima. 

\section{Discussion}
We presented a new construction for a normalized graph Laplacian which generates the backwards Fokker-Planck equation associated to a stochastic differential equation of the form of eq. \ref{eq:driftDiff} with $\mu(x) = -\grad U$ and show how the associated diffusion maps (Fokker-Planck diffusion maps) can be used to represent the dynamics. To do so we first assumed the system is in local equilibrium such that the local density about each point is equivalent to the equilibrium density. By equating the equilibrium density to the kNN density estimate, we derived a simple relationship between the potential, the noise, and the distance to the $k^{th}$ nearest neighbor. If we further assume the noise is equal to a linear function of the distance to the $k^{th}$ nearest neighbor, we can simplify the equation drastically, leading to a simple graph kernel for our system. We then show the differences between our normalization and the most commonly used normalization for diffusion maps introduced in \cite{Haghverdi2016} and used in popular single cell packages like scanpy \cite{Wolf2018}. Finally, we illustrated our graph on single trajectories of stochastic systems with additive and multiplicative noise to show how the diffusion map can capture the dynamics \cite{Nadler2006}, and on an ensemble trajectories in a double well potential localized to a 2-dimensional subspace of the a full 10 dimensional system. We then showed improved hierarchical clustering of the the PBMC3k dataset from illumina. \\

There are many potential directions and improvements to build upon this work, including a more detailed interrogation of the equilibrium condition, the effects of various noise estimates on the diffusion map, and inclusion of anisotropic diffusion. Moreover, since our approach is based on the Fokker-Planck equation, it provides a framework for studying the diffusion limit of scRNAseq data. In particular, it offers a way to connect the ``macroscopic" approach of manifold learning to the more rigorous ``microscopic" approach of chemical master equations as in \cite{Gorin2022, Gorin2023}. These two approaches should coincide in the limit where the number of counts per cell is large, and using Fokker-Planck diffusion maps can help unite these two disparate approaches to the field and put the analysis of scRNAseq on solid physical ground. Additionally, we hope that our approach can offer ways to simulate the dynamics of single cells using approaches such as those laid out in \cite{Coifman2008} and related work.

\bibliography{sde_refs}

\begin{thebibliography}{17}%
\makeatletter
\providecommand \@ifxundefined [1]{%
 \@ifx{#1\undefined}
}%
\providecommand \@ifnum [1]{%
 \ifnum #1\expandafter \@firstoftwo
 \else \expandafter \@secondoftwo
 \fi
}%
\providecommand \@ifx [1]{%
 \ifx #1\expandafter \@firstoftwo
 \else \expandafter \@secondoftwo
 \fi
}%
\providecommand \natexlab [1]{#1}%
\providecommand \enquote  [1]{``#1''}%
\providecommand \bibnamefont  [1]{#1}%
\providecommand \bibfnamefont [1]{#1}%
\providecommand \citenamefont [1]{#1}%
\providecommand \href@noop [0]{\@secondoftwo}%
\providecommand \href [0]{\begingroup \@sanitize@url \@href}%
\providecommand \@href[1]{\@@startlink{#1}\@@href}%
\providecommand \@@href[1]{\endgroup#1\@@endlink}%
\providecommand \@sanitize@url [0]{\catcode `\\12\catcode `\$12\catcode `\&12\catcode `\#12\catcode `\^12\catcode `\_12\catcode `\%12\relax}%
\providecommand \@@startlink[1]{}%
\providecommand \@@endlink[0]{}%
\providecommand \url  [0]{\begingroup\@sanitize@url \@url }%
\providecommand \@url [1]{\endgroup\@href {#1}{\urlprefix }}%
\providecommand \urlprefix  [0]{URL }%
\providecommand \Eprint [0]{\href }%
\providecommand \doibase [0]{https://doi.org/}%
\providecommand \selectlanguage [0]{\@gobble}%
\providecommand \bibinfo  [0]{\@secondoftwo}%
\providecommand \bibfield  [0]{\@secondoftwo}%
\providecommand \translation [1]{[#1]}%
\providecommand \BibitemOpen [0]{}%
\providecommand \bibitemStop [0]{}%
\providecommand \bibitemNoStop [0]{.\EOS\space}%
\providecommand \EOS [0]{\spacefactor3000\relax}%
\providecommand \BibitemShut  [1]{\csname bibitem#1\endcsname}%
\let\auto@bib@innerbib\@empty
%</preamble>
\bibitem [{\citenamefont {Gorin}\ and\ \citenamefont {Pachter}(2022)}]{Gorin2022}%
  \BibitemOpen
  \bibfield  {author} {\bibinfo {author} {\bibfnamefont {G.}~\bibnamefont {Gorin}}\ and\ \bibinfo {author} {\bibfnamefont {L.}~\bibnamefont {Pachter}},\ }\bibfield  {title} {\bibinfo {title} {Distinguishing biophysical stochasticity from technical noise in single-cell rna sequencing using monod}\ }\href {https://doi.org/10.1101/2022.06.11.495771} {10.1101/2022.06.11.495771} (\bibinfo {year} {2022})\BibitemShut {NoStop}%
\bibitem [{\citenamefont {Gorin}\ \emph {et~al.}(2023)\citenamefont {Gorin}, \citenamefont {Vastola},\ and\ \citenamefont {Pachter}}]{Gorin2023}%
  \BibitemOpen
  \bibfield  {author} {\bibinfo {author} {\bibfnamefont {G.}~\bibnamefont {Gorin}}, \bibinfo {author} {\bibfnamefont {J.~J.}\ \bibnamefont {Vastola}},\ and\ \bibinfo {author} {\bibfnamefont {L.}~\bibnamefont {Pachter}},\ }\bibfield  {title} {\bibinfo {title} {Studying stochastic systems biology of the cell with single-cell genomics data}\ }\href {https://doi.org/10.1101/2023.05.17.541250} {10.1101/2023.05.17.541250} (\bibinfo {year} {2023})\BibitemShut {NoStop}%
\bibitem [{\citenamefont {Coifman}\ and\ \citenamefont {Lafon}(2006)}]{Coifman2006}%
  \BibitemOpen
  \bibfield  {author} {\bibinfo {author} {\bibfnamefont {R.~R.}\ \bibnamefont {Coifman}}\ and\ \bibinfo {author} {\bibfnamefont {S.}~\bibnamefont {Lafon}},\ }\bibfield  {title} {\bibinfo {title} {Diffusion maps},\ }\href {https://doi.org/10.1016/j.acha.2006.04.006} {\bibfield  {journal} {\bibinfo  {journal} {Applied and Computational Harmonic Analysis}\ }\textbf {\bibinfo {volume} {21}},\ \bibinfo {pages} {5} (\bibinfo {year} {2006})}\BibitemShut {NoStop}%
\bibitem [{\citenamefont {Coifman}\ \emph {et~al.}(2005)\citenamefont {Coifman}, \citenamefont {Lafon}, \citenamefont {Lee}, \citenamefont {Maggioni}, \citenamefont {Nadler}, \citenamefont {Warner},\ and\ \citenamefont {Zucker}}]{Coifman2005}%
  \BibitemOpen
  \bibfield  {author} {\bibinfo {author} {\bibfnamefont {R.~R.}\ \bibnamefont {Coifman}}, \bibinfo {author} {\bibfnamefont {S.}~\bibnamefont {Lafon}}, \bibinfo {author} {\bibfnamefont {A.~B.}\ \bibnamefont {Lee}}, \bibinfo {author} {\bibfnamefont {M.}~\bibnamefont {Maggioni}}, \bibinfo {author} {\bibfnamefont {B.}~\bibnamefont {Nadler}}, \bibinfo {author} {\bibfnamefont {F.}~\bibnamefont {Warner}},\ and\ \bibinfo {author} {\bibfnamefont {S.~W.}\ \bibnamefont {Zucker}},\ }\bibfield  {title} {\bibinfo {title} {Geometric diffusions as a tool for harmonic analysis and structure definition of data: Diffusion maps},\ }\href {https://doi.org/10.1073/pnas.0500334102} {\bibfield  {journal} {\bibinfo  {journal} {Proceedings of the National Academy of Sciences}\ }\textbf {\bibinfo {volume} {102}},\ \bibinfo {pages} {7426} (\bibinfo {year} {2005})}\BibitemShut {NoStop}%
\bibitem [{\citenamefont {Ting}\ \emph {et~al.}(2011)\citenamefont {Ting}, \citenamefont {Huang},\ and\ \citenamefont {Jordan}}]{Ting2011}%
  \BibitemOpen
  \bibfield  {author} {\bibinfo {author} {\bibfnamefont {D.}~\bibnamefont {Ting}}, \bibinfo {author} {\bibfnamefont {L.}~\bibnamefont {Huang}},\ and\ \bibinfo {author} {\bibfnamefont {M.}~\bibnamefont {Jordan}},\ }\bibfield  {title} {\bibinfo {title} {An analysis of the convergence of graph laplacians}\ }\href {https://doi.org/10.48550/ARXIV.1101.5435} {10.48550/ARXIV.1101.5435} (\bibinfo {year} {2011})\BibitemShut {NoStop}%
\bibitem [{\citenamefont {Huang}(2011)}]{Huang2011}%
  \BibitemOpen
  \bibfield  {author} {\bibinfo {author} {\bibfnamefont {S.}~\bibnamefont {Huang}},\ }\bibfield  {title} {\bibinfo {title} {The molecular and mathematical basis of waddington{\textquotesingle}s epigenetic landscape: A framework for post-darwinian biology?},\ }\href {https://doi.org/10.1002/bies.201100031} {\bibfield  {journal} {\bibinfo  {journal} {{BioEssays}}\ }\textbf {\bibinfo {volume} {34}},\ \bibinfo {pages} {149} (\bibinfo {year} {2011})}\BibitemShut {NoStop}%
\bibitem [{\citenamefont {Wang}\ \emph {et~al.}(2011)\citenamefont {Wang}, \citenamefont {Zhang}, \citenamefont {Xu},\ and\ \citenamefont {Wang}}]{wang_quantifying_2011}%
  \BibitemOpen
  \bibfield  {author} {\bibinfo {author} {\bibfnamefont {J.}~\bibnamefont {Wang}}, \bibinfo {author} {\bibfnamefont {K.}~\bibnamefont {Zhang}}, \bibinfo {author} {\bibfnamefont {L.}~\bibnamefont {Xu}},\ and\ \bibinfo {author} {\bibfnamefont {E.}~\bibnamefont {Wang}},\ }\bibfield  {title} {\bibinfo {title} {Quantifying the {Waddington} landscape and biological paths for development and differentiation},\ }\href {https://doi.org/10.1073/pnas.1017017108} {\bibfield  {journal} {\bibinfo  {journal} {Proceedings of the National Academy of Sciences}\ }\textbf {\bibinfo {volume} {108}},\ \bibinfo {pages} {8257} (\bibinfo {year} {2011})},\ \bibinfo {note} {publisher: Proceedings of the National Academy of Sciences}\BibitemShut {NoStop}%
\bibitem [{\citenamefont {Ferrell}(2012)}]{Ferrell2012}%
  \BibitemOpen
  \bibfield  {author} {\bibinfo {author} {\bibfnamefont {J.~E.}\ \bibnamefont {Ferrell}},\ }\bibfield  {title} {\bibinfo {title} {Bistability, bifurcations, and waddington{\textquotesingle}s epigenetic landscape},\ }\href {https://doi.org/10.1016/j.cub.2012.03.045} {\bibfield  {journal} {\bibinfo  {journal} {Current Biology}\ }\textbf {\bibinfo {volume} {22}},\ \bibinfo {pages} {R458} (\bibinfo {year} {2012})}\BibitemShut {NoStop}%
\bibitem [{\citenamefont {Haghverdi}\ \emph {et~al.}(2015)\citenamefont {Haghverdi}, \citenamefont {Buettner},\ and\ \citenamefont {Theis}}]{Haghverdi2015}%
  \BibitemOpen
  \bibfield  {author} {\bibinfo {author} {\bibfnamefont {L.}~\bibnamefont {Haghverdi}}, \bibinfo {author} {\bibfnamefont {F.}~\bibnamefont {Buettner}},\ and\ \bibinfo {author} {\bibfnamefont {F.~J.}\ \bibnamefont {Theis}},\ }\bibfield  {title} {\bibinfo {title} {Diffusion maps for high-dimensional single-cell analysis of differentiation data},\ }\href {https://doi.org/10.1093/bioinformatics/btv325} {\bibfield  {journal} {\bibinfo  {journal} {Bioinformatics}\ }\textbf {\bibinfo {volume} {31}},\ \bibinfo {pages} {2989} (\bibinfo {year} {2015})}\BibitemShut {NoStop}%
\bibitem [{\citenamefont {Haghverdi}\ \emph {et~al.}(2016)\citenamefont {Haghverdi}, \citenamefont {B\"{u}ttner}, \citenamefont {Wolf}, \citenamefont {Buettner},\ and\ \citenamefont {Theis}}]{Haghverdi2016}%
  \BibitemOpen
  \bibfield  {author} {\bibinfo {author} {\bibfnamefont {L.}~\bibnamefont {Haghverdi}}, \bibinfo {author} {\bibfnamefont {M.}~\bibnamefont {B\"{u}ttner}}, \bibinfo {author} {\bibfnamefont {F.~A.}\ \bibnamefont {Wolf}}, \bibinfo {author} {\bibfnamefont {F.}~\bibnamefont {Buettner}},\ and\ \bibinfo {author} {\bibfnamefont {F.~J.}\ \bibnamefont {Theis}},\ }\bibfield  {title} {\bibinfo {title} {Diffusion pseudotime robustly reconstructs lineage branching},\ }\href {https://doi.org/10.1038/nmeth.3971} {\bibfield  {journal} {\bibinfo  {journal} {Nature Methods}\ }\textbf {\bibinfo {volume} {13}},\ \bibinfo {pages} {845} (\bibinfo {year} {2016})}\BibitemShut {NoStop}%
\bibitem [{\citenamefont {Wolf}\ \emph {et~al.}(2018)\citenamefont {Wolf}, \citenamefont {Angerer},\ and\ \citenamefont {Theis}}]{Wolf2018}%
  \BibitemOpen
  \bibfield  {author} {\bibinfo {author} {\bibfnamefont {F.~A.}\ \bibnamefont {Wolf}}, \bibinfo {author} {\bibfnamefont {P.}~\bibnamefont {Angerer}},\ and\ \bibinfo {author} {\bibfnamefont {F.~J.}\ \bibnamefont {Theis}},\ }\bibfield  {title} {\bibinfo {title} {{SCANPY}: large-scale single-cell gene expression data analysis},\ }\bibfield  {journal} {\bibinfo  {journal} {Genome Biology}\ }\textbf {\bibinfo {volume} {19}},\ \href {https://doi.org/10.1186/s13059-017-1382-0} {10.1186/s13059-017-1382-0} (\bibinfo {year} {2018})\BibitemShut {NoStop}%
\bibitem [{\citenamefont {Stuart}\ \emph {et~al.}(2019)\citenamefont {Stuart}, \citenamefont {Butler}, \citenamefont {Hoffman}, \citenamefont {Hafemeister}, \citenamefont {Papalexi}, \citenamefont {Mauck}, \citenamefont {Hao}, \citenamefont {Stoeckius}, \citenamefont {Smibert},\ and\ \citenamefont {Satija}}]{Stuart2019}%
  \BibitemOpen
  \bibfield  {author} {\bibinfo {author} {\bibfnamefont {T.}~\bibnamefont {Stuart}}, \bibinfo {author} {\bibfnamefont {A.}~\bibnamefont {Butler}}, \bibinfo {author} {\bibfnamefont {P.}~\bibnamefont {Hoffman}}, \bibinfo {author} {\bibfnamefont {C.}~\bibnamefont {Hafemeister}}, \bibinfo {author} {\bibfnamefont {E.}~\bibnamefont {Papalexi}}, \bibinfo {author} {\bibfnamefont {W.~M.}\ \bibnamefont {Mauck}}, \bibinfo {author} {\bibfnamefont {Y.}~\bibnamefont {Hao}}, \bibinfo {author} {\bibfnamefont {M.}~\bibnamefont {Stoeckius}}, \bibinfo {author} {\bibfnamefont {P.}~\bibnamefont {Smibert}},\ and\ \bibinfo {author} {\bibfnamefont {R.}~\bibnamefont {Satija}},\ }\bibfield  {title} {\bibinfo {title} {Comprehensive integration of single-cell data},\ }\href {https://doi.org/10.1016/j.cell.2019.05.031} {\bibfield  {journal} {\bibinfo  {journal} {Cell}\ }\textbf {\bibinfo {volume} {177}},\ \bibinfo {pages} {1888} (\bibinfo {year} {2019})}\BibitemShut {NoStop}%
\bibitem [{\citenamefont {Coifman}\ \emph {et~al.}(2008)\citenamefont {Coifman}, \citenamefont {Kevrekidis}, \citenamefont {Lafon}, \citenamefont {Maggioni},\ and\ \citenamefont {Nadler}}]{Coifman2008}%
  \BibitemOpen
  \bibfield  {author} {\bibinfo {author} {\bibfnamefont {R.~R.}\ \bibnamefont {Coifman}}, \bibinfo {author} {\bibfnamefont {I.~G.}\ \bibnamefont {Kevrekidis}}, \bibinfo {author} {\bibfnamefont {S.}~\bibnamefont {Lafon}}, \bibinfo {author} {\bibfnamefont {M.}~\bibnamefont {Maggioni}},\ and\ \bibinfo {author} {\bibfnamefont {B.}~\bibnamefont {Nadler}},\ }\bibfield  {title} {\bibinfo {title} {Diffusion maps, reduction coordinates, and low dimensional representation of stochastic systems},\ }\href {https://doi.org/10.1137/070696325} {\bibfield  {journal} {\bibinfo  {journal} {Multiscale Modeling and Simulation}\ }\textbf {\bibinfo {volume} {7}},\ \bibinfo {pages} {842} (\bibinfo {year} {2008})}\BibitemShut {NoStop}%
\bibitem [{\citenamefont {van Kampen}(1981)}]{vanKampen1981}%
  \BibitemOpen
  \bibfield  {author} {\bibinfo {author} {\bibfnamefont {N.~G.}\ \bibnamefont {van Kampen}},\ }\bibfield  {title} {\bibinfo {title} {It{\^{o}} versus stratonovich},\ }\href {https://doi.org/10.1007/bf01007642} {\bibfield  {journal} {\bibinfo  {journal} {Journal of Statistical Physics}\ }\textbf {\bibinfo {volume} {24}},\ \bibinfo {pages} {175} (\bibinfo {year} {1981})}\BibitemShut {NoStop}%
\bibitem [{\citenamefont {Coomer}\ \emph {et~al.}(2022)\citenamefont {Coomer}, \citenamefont {Ham},\ and\ \citenamefont {Stumpf}}]{Coomer2022}%
  \BibitemOpen
  \bibfield  {author} {\bibinfo {author} {\bibfnamefont {M.~A.}\ \bibnamefont {Coomer}}, \bibinfo {author} {\bibfnamefont {L.}~\bibnamefont {Ham}},\ and\ \bibinfo {author} {\bibfnamefont {M.~P.}\ \bibnamefont {Stumpf}},\ }\bibfield  {title} {\bibinfo {title} {Noise distorts the epigenetic landscape and shapes cell-fate decisions},\ }\href {https://doi.org/10.1016/j.cels.2021.09.002} {\bibfield  {journal} {\bibinfo  {journal} {Cell Systems}\ }\textbf {\bibinfo {volume} {13}},\ \bibinfo {pages} {83} (\bibinfo {year} {2022})}\BibitemShut {NoStop}%
\bibitem [{\citenamefont {Loftsgaarden}\ and\ \citenamefont {Quesenberry}(1965)}]{Loftsgaarden1965}%
  \BibitemOpen
  \bibfield  {author} {\bibinfo {author} {\bibfnamefont {D.~O.}\ \bibnamefont {Loftsgaarden}}\ and\ \bibinfo {author} {\bibfnamefont {C.~P.}\ \bibnamefont {Quesenberry}},\ }\bibfield  {title} {\bibinfo {title} {A nonparametric estimate of a multivariate density function},\ }\href {https://doi.org/10.1214/aoms/1177700079} {\bibfield  {journal} {\bibinfo  {journal} {The Annals of Mathematical Statistics}\ }\textbf {\bibinfo {volume} {36}},\ \bibinfo {pages} {1049–1051} (\bibinfo {year} {1965})}\BibitemShut {NoStop}%
\bibitem [{\citenamefont {Nadler}\ \emph {et~al.}(2006)\citenamefont {Nadler}, \citenamefont {Lafon}, \citenamefont {Coifman},\ and\ \citenamefont {Kevrekidis}}]{Nadler2006}%
  \BibitemOpen
  \bibfield  {author} {\bibinfo {author} {\bibfnamefont {B.}~\bibnamefont {Nadler}}, \bibinfo {author} {\bibfnamefont {S.}~\bibnamefont {Lafon}}, \bibinfo {author} {\bibfnamefont {R.~R.}\ \bibnamefont {Coifman}},\ and\ \bibinfo {author} {\bibfnamefont {I.~G.}\ \bibnamefont {Kevrekidis}},\ }\bibfield  {title} {\bibinfo {title} {Diffusion maps, spectral clustering and reaction coordinates of dynamical systems},\ }\href {https://doi.org/10.1016/j.acha.2005.07.004} {\bibfield  {journal} {\bibinfo  {journal} {Applied and Computational Harmonic Analysis}\ }\textbf {\bibinfo {volume} {21}},\ \bibinfo {pages} {113} (\bibinfo {year} {2006})}\BibitemShut {NoStop}%
\end{thebibliography}%

\end{document}